\documentclass[aps,prd,preprint,groupedaddress,longbibliography]{revtex4-1}

\usepackage{amsmath,amssymb,amsfonts}
\usepackage[T1]{fontenc}
\usepackage{graphicx}
\usepackage{array}
\newcolumntype{L}{>{\centering\arraybackslash}p{3cm}}
\usepackage{hyperref}

\begin{document}

\title{Neutron-mirror neutron oscillations for solving the puzzles of ultrahigh-energy cosmic rays}

\author{Wanpeng Tan}
\email[]{wtan@nd.edu}
\affiliation{Department of Physics, Institute for Structure and Nuclear Astrophysics (ISNAP), and Joint Institute for Nuclear Astrophysics - Center for the Evolution of Elements (JINA-CEE), University of Notre Dame, Notre Dame, Indiana 46556, USA}

\date{\today}

\begin{abstract}
Based on a newly proposed mirror-matter model of neutron-mirror neutron ($n-n'$) oscillations, the puzzles related to ultrahigh-energy cosmic rays (UHECRs) are explained. In particular, the phenomena around the Greisen-Zatsepin-Kuzmin (GZK) cutoff for UHECRs can be well understood under the new mirror matter model assuming a mirror-to-ordinary temperature ratio of $T'/T \sim 0.3$. The suppression factor of the GZK effect due to the opacity of cosmic microwave background is calculated and agrees with the observations well. Most of the super-GZK events (i.e., above the GZK cutoff), as predicted in the new model, come from mirror matter sources that are invisible to  electromagnetic telescopes and can penetrate the mirror cosmic microwave background at much further distances. Most remarkably, the anti-correlation between super-GZK and sub-GZK events in the hotspot observed by the Telescope Array (TA) collaboration can be naturally understood in this model. The possible correlations between the UHECRs from the TA hotspot and other nearby powerful sources such as high energy neutrinos detected by IceCube, the largest black hole merger (GW170729) observed by LIGO, and the hottest star-forming supercluster Lynx Arc, are discussed as well under the new theory.
\end{abstract}


\maketitle

Ultrahigh-energy cosmic rays (UHECRs) with energies up to the order of $10^{20}$ eV have become a window for search of new physics beyond the standard model as some of these high energy particles are millions of times more energetic than anything we can produce at the most powerful accelerator on the planet (see Refs. \cite{anchordoqui2019,alvesbatista2019} for recent reviews on UHECRs). The UHECR protons can lose energy by interacting with relic photons of cosmic microwave background (CMB) via the following photopion reactions,
\begin{equation}\label{eq_pg}
p + \gamma \rightarrow n + \pi^+ ( p + \pi^0 )
\end{equation}
where the threshold energy can be estimated from 
\begin{equation}\label{eq_GZK}
E_{p} \geq \frac{m_{\pi}m_p+m^2_{\pi}/2}{E_{\gamma}(1-\cos \theta)} \geq \frac{m_{\pi}m_p+m^2_{\pi}/2}{2E_{\gamma}} \equiv E_{GZK} \sim 6\times 10^{19} \text{eV}
\end{equation}
for the typical relic photon energy $E_{\gamma} = 10^{-3}$ eV corresponding to the CMB temperature of 2.73 K. On average, protons (or neutrons similarly) with energy above $E_{GZK}$ lose about 20\% energy per scattering and the mean free path of the photopion reactions can be estimated by $l_{\pi} = 1/(\sigma_{\pi} n_{\gamma}) \sim 10^7$ lightyears where the photopion cross section $\sigma_{\pi} \sim 0.2$ mb and relic photon number density $n_{\gamma} = 500$ cm$^{-3}$. Therefore, UHECR protons will quickly lose most of their energy to stay below $E_{GZK}$ when propagating at large cosmological distances and being scattered up to $10^3$ times. The cosmic ray spectrum will show an abrupt cutoff at the so-called Greisen-Zatsepin-Kuzmin (GZK) limit $E_{GZK}$ due to the opacity of CMB \cite{greisen1966,zatsepin1966}. Indeed, such a suppression at energies of $4-6\times 10^{19}$ eV was observed by large UHECR detector arrays \cite{thepierreaugercollaboration2008,highresolutionflyseyecollaboration2008}.

However, there are quite a few puzzles for the UHECRs. One of the main problems is where the UHECRs come from. As widely accepted, UHECRs, at least with energies above $10^{19}$ eV, have an extra-galactic origin, which was essentially confirmed by anisotropy observations by the Pierre Auger and Telescope Array (TA) collaborations \cite{abbasi2014,thepierreaugercollaboration2017,abbasi2018}. These UHECR particles are nearly unaffected by galactic and intergalactic magnetic fields and the mapping of their directionality should help identify the location of the sources. But there have been observations of super-GZK events with energies above $10^{20}$ eV that are difficult to explain under the GZK effect and in addition, no obvious sources can be connected to these events within the CMB attenuated distance \cite{hayashida1994,berezinsky2006,abbasi2018}. 
UHECR mass composition is typically inferred from observations of the extensive air showers, i.e., the depth where the shower reaches its maximum, and depends on the models of hadronic interactions at extreme energies. Heavier composition at higher energies was used to fit the energy spectrum of UHECRs \cite{anchordoqui2019,alvesbatista2019}. However, new physics beyond the standard model at such high energies could make UHECR protons appear heavier under the normal interpretation \cite{farrar2013,pavlidou2019}. New insights in hadron physics rather than the heavy nuclei assumption are needed for us to fully understand the air shower physics including the muon excess puzzle. In particular, Farrar and Allen \cite{farrar2013} suggested a chiral symmetry restoration model that can provide a remarkably consistent description of air shower observations with UHECRs as pure protons and is in line with the idea of staged quark condensation \cite{tan2019c}. In addition, nucleosynthesis from big bang and stars indicates that hydrogen is much more abundant than other heavy elements and more so in the early Universe. Therefore, protons should be a natural dominant source for UHECRs.

An even more puzzling observation by the TA collaboration \cite{abbasi2018} shows that in the hotspot direction there is a deficit of sub-GZK events with energies in between $10^{19.2}$ and $10^{19.75}$ eV and
an excess for super-GZK events ($E > 10^{19.75}$ eV), which seems to indicate that sub-GZK and super-GZK events have different origins. On the other hand, the UHECRs from the hotspot observed by the TA collaboration are likely correlated with IceCube neutrinos \cite{fang2014,icecube2016}, which may indicate a single source origin for both UHECRs and neutrinos.

In the following, the above-discussed puzzles of UHECRs will be explained naturally under the newly developed $n-n'$ oscillation model \cite{tan2019}. It is based on the mirror matter theory \cite{kobzarev1966,blinnikov1982,blinnikov1983,kolb1985,khlopov1991,foot2004,berezhiani2004,okun2007,cui2012}, that is, two sectors of particles have similar yet separate gauge
interactions within their own sector but share the same gravitational force. Such a mirror matter theory has appealing theoretical features. For example, it can be embedded in the $E_8\otimes E_{8'}$ superstring theory \cite{green1984,gross1985,kolb1985} and it can also be a natural extension of recently developed twin Higgs models \cite{chacko2006,barbieri2005} that protect the Higgs mass from quadratic divergences and hence solve the hierarchy or fine-tuning problem. The mirror symmetry or twin Higgs mechanism is particularly intriguing as the Large Hadron Collider has found no evidence of supersymmetry so far and we may not need supersymmetry, at least not below energies of 10 TeV. The new mirror matter model can consistently and quantitatively explain various intriguing phenomena and observations in the universe including the neutron lifetime puzzle and dark-to-baryon matter ratio \cite{tan2019}, evolution and nucleosynthesis in stars \cite{tan2019a}, matter-antimatter asymmetry of the universe \cite{tan2019c}, unitarity of the CKM matrix \cite{tan2019d}, invisible decays of neutral hadrons \cite{tan2020d}, and dark energy \cite{tan2019e}. Further development of the model into a dynamic supersymmetric mirror theory can be seen in Refs. \cite{tan2020,tan2020a} and applied to understand the nature of black holes \cite{tan2020b}.

In this new mirror matter model \cite{tan2019}, no cross-sector interaction is introduced, unlike other $n-n'$ type models. The critical assumption of this model is that the mirror symmetry is spontaneously broken by the uneven Higgs vacuum in the two sectors, i.e., $<\phi> \neq <\phi'>$, although very slightly (on a relative breaking scale of $10^{-15} \text{--} 10^{-14}$) \cite{tan2019}. When fermion particles obtain their mass from the Yukawa coupling, it automatically leads to the mirror mixing for neutral particles, i.e., the basis of mass eigenstates is not the same as that of mirror eigenstates, similar to the case of ordinary neutrino oscillations due to the family or generation mixing. Further details of the model can be found in Ref. \cite{tan2019}.

The immediate result of this model is the probability of $n-n'$ oscillations in vacuum \cite{tan2019},
\begin{equation}\label{eq_prob}
P_{nn'}(t) = \sin^2(2\theta) \sin^2(\frac{1}{2}\Delta_{nn'} t)
\end{equation}
where $\theta$ is the $n-n'$ mixing angle and $\sin^2(2\theta)$ denotes the mixing strength of about $2\times 10^{-5}$, $t$ is the propagation time, $\Delta_{nn'} = m_{n2} - m_{n1}$ is the small mass difference of the two mass eigenstates of about $2\times 10^{-6}$ eV \cite{tan2019}, and natural units ($\hbar=c=1$) are used for simplicity. Note that the equation is valid even for relativistic neutrons and in this case $t$ is the proper time in the particle's rest frame. The two model parameters of $\sin^2(2\theta)$ and $\Delta_{nn'}$ are fairly well constrained by neutron lifetime measurements and observation of dark matter and baryon asymmetry \cite{tan2019,tan2019c}. In this work, we will present an estimate on the cosmology parameter of the model, i.e., the CMB temperature ratio of the two sectors.

For neutrons travel in the CMB medium, each collision or interaction with a relic photon will collapse the oscillating wave function into a mirror eigenstate, in other words, during mean free flight time $\tau_f$ the $n-n'$ transition probability is $P_{nn'}(\tau_f)$. The number of such collisions will be $1/\tau_f$ in a unit time. Therefore, the transition rate of $n-n'$ for in-medium neutrons is \cite{tan2019},
\begin{equation}\label{eq_prob2}
\lambda_{nn'} = \frac{1}{\tau_f}\sin^2(2\theta) <\sin^2(\frac{1}{2}\Delta_{nn'} \tau_f)>.
\end{equation}
Note that the Mikheyev-Smirnov-Wolfenstein (MSW) matter effect \cite{wolfenstein1978,mikheev1985}, i.e., coherent forward scattering that could affect the oscillations is negligible as the neutron-photon scattering cross section is very small \cite{gould1993} and the CMB photon density is too low (see more details for in-medium $n-n'$ oscillations from Ref. \cite{tan2019a}).

Explanation of excess of super-GZK protons was attempted under the consideration of $n-n'$ oscillations with a different mechanism as discussed in Ref. \cite{berezhiani2006a} with a caveat of unrealistic constraints on galactic and intergalactic magnetic fields. Under the new $n-n'$ oscillation model, the situation is different. For super-GZK protons around energy of $E_p=2\times 10^{20}$ eV, the Lorentz factor $\gamma = E_p/m_p \sim 2\times 10^{11}$ and the pair production reaction ($p\gamma \rightarrow pe^-e^+$) has a higher cross section of about 10 mb \cite{hart1959} leading to the corresponding mean free path of $l_{pair} \sim 2\times10^5$ lightyears in the ordinary world. Note that $l_{pair}$ is much shorter than $l_{\pi}$ and we will demonstrate below how this can affect $n-n'$ oscillations and alleviate the suppression of super-GZK events to agree with the observations.

Both ordinary and mirror sectors have almost identical micro-physics and parameters except the mirror world may have a much lower temperature $T'$ than the ordinary world temperature $T$ \cite{kolb1985,hodges1993,foot2004,berezhiani2004,tan2019,tan2019c}.
To be consistent with the results of the standard big bang nucleosynthesis (BBN) model, in particular, the well known primordial helium abundance, a strict requirement of $T'/T < 1/2$ at BBN temperatures \cite{kolb1985,hodges1993,foot2004,berezhiani2004} has to be met to ensure a slow enough expansion of the universe. Such a temperature condition can naturally occur after the early inflation and subsequent reheating \cite{kolb1985,hodges1993}.

For the mirror world with a lower CMB temperature that meets the above requirement, for example, a typical ratio of $x=T'/T \sim 0.3$ \cite{berezhiani2006a}, the mean free path of UHECR mirror protons will be $l' = x^{-3} l \sim 40l$ for both photopion and $e^-e^+$ pair production reactions. The $p\gamma$ and $n\gamma$ cross sections should be similar for photopion and pair production reactions, respectively. Therefore, 
the mean free flight time for a super-GZK neutron in its rest frame $\tau_f = l_{pair}/(\gamma c) \sim 20$ s and the corresponding mean free flight time in the mirror world would be $\tau'_f = l'_{pair}/(\gamma c) \sim 800$ s. Then we can estimate the fraction of the successful $n-n'$ transition in the ordinary world,
\begin{equation}\label{eq_probf}
f = \frac{\tau_{\beta}}{\tau_f}\sin^2(2\theta) <\sin^2(\frac{1}{2}\Delta_{nn'} \tau_{\beta})> = \frac{\tau_{\beta}}{\tau_f} 10^{-5} \sim 5\times 10^{-4}
\end{equation}
where $\tau_{\beta} = 888$ s is the neutron $\beta$-decay lifetime \cite{tan2019} and $f' = 10^{-5}$ can be obtained similarly for the mirror world.

We first consider the $n-n'$ oscillations from the ordinary sector. A super-GZK proton is first converted to a neutron by photopion reaction that then oscillate into a mirror neutron. The mirror neutron/proton can travel in the mirror CMB medium $x^{-3} \sim 40$ times as far as the ordinary one. Before it arrives on Earth, it oscillates back into an ordinary neutron. In this two-step oscillation scenario, the overall suppression factor for super-GZK events would be $f\times f'=5\times 10^{-9}$ which is too small to be observed.

On the contrary, the super-GZK events could be from the mirror world. In this case, A super-GZK mirror proton travels in the mirror CMB medium up to 40 times as far as the ordinary one in the ordinary world. When it is close to the Earth, it is converted to a mirror neutron by mirror photopion reaction that is then oscillated into an ordinary neutron. Since the mirror world is about 5.4 times as dense as the ordinary world inferred from the observed dark-to-baryon matter density ratio \cite{tan2019} and the mean free path is 40 times as large, we can obtain an enhancement factor for the super-GZK events of about 200. Under this scenario, therefore, the overall suppression factor for super-GZK events would be $200\times f' = 2\times10^{-3}$ which is very similar to the suppression factor of the UHECR flux $J(E)$ at $E = 2\times 10^{20}$ eV due to the GZK cutoff observed by the Auger and TA collaborations \cite{abu-zayyad2013,verzi2017}.

The new $n-n'$ oscillation model \cite{tan2019}, as discussed above, essentially predicts that most of the observed UHECR protons above $E_{GZK}$ come from the mirror sources. As the mirror objects can not be directly observed via electromagnetic radiation, this naturally explains why the super-GZK events could not be connected with any ordinary sources \cite{hayashida1994,berezinsky2006,abbasi2018}. In addition, mirror UHECR protons with energies below $E_{GZK}$ from the same source have no chance to transition to ordinary ones as the energy is below the threshold of mirror photopion reaction. Accordingly the observed super-GZK events appear alone without lower energy companions, which explains the observed anti-correlation between super-GZK and sub-GZK events by the TA collaboration \cite{abbasi2018}. As discussed above, the super-GZK events stem mainly from the mirror UHECR sources while the sub-GZK events are dominated by the ordinary sources. As confirmed by observations, these UHECRs are of extra-galactic origin and anisotropic in the arrival directions \cite{abbasi2014,thepierreaugercollaboration2017,abbasi2018}. The deficit of sub-GZK events in the TA super-GZK hotspot, therefore, means that strong mirror sources and less ordinary sources are aligned in this direction. Under the mirror-matter theory, such inhomogeneities are typically generated in the early universe during the structure formation and segregation of ordinary and mirror matter at large scales \cite{blinnikov1983,kolb1985}.

In the new mirror matter model \cite{tan2019}, not only neutrons but other neutral particles oscillate as well. In particular, neutrino-mirror neutrino oscillations will follow the transition probability,
\begin{equation}\label{eq_probnu}
P_{\nu \nu'}(t) = \sin^2(2\theta_{\nu\nu'}) \sin^2(\frac{\Delta^2_{\nu\nu'}}{4E} t)
\end{equation}
where the mass difference $\Delta^2_{\nu\nu'} \sim 10^{-18}$ eV$^2$ is much smaller than the mass difference in normal neutrino flavor mixing \cite{tan2019}, $E$ is the neutrino energy, and $t$ is the propagation time in the Earth frame. The mixing angle $\theta_{\nu\nu'}$ should be similar to the values in neutrino flavor oscillations and hence the mixing strength should be fairly large, i.e., $ \sin^2(2\theta_{\nu\nu'}) \gtrsim 0.1$. To make the propagation factor averaged to $1/2$ for neutrinos of $E \sim 10^{14}$ eV observed by the IceCube collaboration \cite{fang2014,icecube2016}, the source distance has to be larger than $3\times 10^9$ lightyears.

For super-GZK events from the ordinary world, the source distance has to be constrained within $3\times 10^8$ lightyears due to the GZK effect \cite{aharonian1994}. However, that distance can be relaxed to $10^{10}$ lightyears or more (close to the size of our visible Universe) for the mirror sources depending on the parameter of $x=T'/T$. Therefore, if a mirror source accounted for the super-GZK events emits mirror neutrinos of $E = 10^{14}$ eV at the same time, a significant portion of these mirror neutrinos will oscillate into ordinary neutrinos and be detected at the Earth by neutrino detectors in coincidence with the UHECR protons. Even if the vacuum mixing strength is low, such high energy mirror neutrinos could experience resonant oscillations due to the mirror MSW matter effect ending up with more ordinary neutrinos \cite{mikheev1985}. The observed correlation between the super-GZK events from the TA hotspot and neutrinos detected by the IceCube observatory \cite{fang2014,icecube2016} shows evidence of such a possible mirror source.

The TA super-GZK hotspot (about $20^{\circ}$ in size) at $9^{\text{h}} 16^{\text{m}}, 45^{\circ}$ \cite{abbasi2018} could be located in the same direction of the largest black hole merger (GW170729) observed by LIGO \cite{theligoscientificcollaboration2018}. No optical counterpart has been identified for these black hole mergers which could very well reside in mirror matter dominated regions. In addition, the luminosity distance of GW170729 is estimated at $2750\pm 1350$ Mpc \cite{theligoscientificcollaboration2018} which is consistent with the estimate for the super-GZK events and the requirement for the neutrino-UHECR coincidences discussed above. In a similar direction ($8^{\text{h}} 48^{\text{m}}, 44^{\circ}55'$), the hottest and largest star-forming region Lynx Arc was also found at a comparable distance of about $1.2\times 10^{10}$ ly with gravitational lensing technique \cite{fosbury2003}. This TA hotspot may have started to reveal one of the most active and powerful mirror matter sources in the early universe and it certainly deserves further and more detailed studies.

Such super-GZK events from much more distant mirror sources rather than closer ordinary ones may indicate that the most powerful acceleration sites for UHECRs are at the early stages of (mirror) star and galaxy formation in the early universe. So are the sites for the most massive black hole mergers and the hottest star-forming regions. Even if these sites in the early universe produce ordinary super-GZK protons, we can not observe them on Earth today due to the large cosmological distance and the opacity of CMB. The ones created by nearby sites at the early time are long gone from our local region. For this reason most of the observed super-GZK events should come from mirror sources at far distances on the order of $10^{10}$ ly. We need more UHECR observatories, in particular, the ones in space \cite{casolino2017} to discover more of these super-GZK sources.

To conclude, the new mirror-matter theory provides a natural explanation for various puzzles related to UHECRs assuming a mirror-to-ordinary CMB temperature ratio of $T'/T \sim 0.3$. The super-GZK events reveal some of the most energetic mirror matter sources in the universe and may one day present clues on how super-GZK particles are accelerated. In the era of multimessenger astrophysics other types of detection via gravitational waves and neutrinos correlated to the super-GZK events are necessary for further understanding of the mirror sources. In particular, more data for correlation between super-GZK hotspots and high energy neutrinos should be pursued. Gravitational lensing studies near the super-GZK hotspots may show us more about the mirror sources and provide further test of the new mirror-matter model.

\begin{acknowledgments}
I would like to thank Jiangping Hu, Hui Hua, and Xiaodong Tang for their hospitality as part of this work was prepared during my visit at their institutions.
This work is supported in part by the National Science Foundation under
grant No. PHY-1713857 and the Joint Institute for Nuclear Astrophysics (JINA-CEE, www.jinaweb.org), NSF-PFC under grant No. PHY-1430152.
\end{acknowledgments}

\bibliography{uhecr}

\end{document}